\documentclass[a4paper,11pt,twoside]{book}
\usepackage[left=1in,right=1in,bottom=1in,top=1in]{geometry}
\usepackage[T1]{fontenc}
\usepackage[utf8]{inputenc}
\usepackage{imakeidx}
\usepackage{graphicx}
\usepackage{lmodern}
\usepackage{hyperref}
\usepackage{listings}
\usepackage{xcolor}
\usepackage{amsmath}
\usepackage{fancyhdr}
\usepackage{emptypage}
\lstdefinestyle{customstyle}{
  belowcaptionskip=1\baselineskip,
  breaklines=true,
  xleftmargin=\parindent,
  language=Python,
  showstringspaces=false,
  basicstyle=\footnotesize\ttfamily,
  keywordstyle=\bfseries\color{green!40!black},
  commentstyle=\itshape\color{purple!40!black},
  identifierstyle=\color{blue},
  stringstyle=\color{orange},
}
\makeindex[options= -s indexstyle.ist]
\title{\textbf{\LARGE Urban Delay Tolerant Network
    Simulator\\(UDTNSim~v0.1)}\vspace{-7mm}} \author{\vspace{4mm}Sarath
  Babu, Gaurav Jain, and B. S. Manoj\\Department of
  Avionics\\
  Indian Institute of Space Science and Technology\\\vspace{2mm}
  Thiruvananthapuram, India 695547\\\vspace{2mm}\small Email: \tt
  sarath.babu.2014@ieee.org, tashan1492@yahoo.com,
  bsmanoj@ieee.org\\\small UDTNSim Source:
  \href{https://github.com/iist-sysnet/UDTNSim}{\tt
    https://github.com/iist-sysnet/UDTNSim}}

\begin{document}
\pagenumbering{roman}
\maketitle

\cleardoublepage
\section*{\begin{center}Abstract\end{center}}
Delay Tolerant Networking (DTN) is an approach to networking which
handles network disruptions and high delays that may occur in many
kinds of communication networks. The major reasons for high delay
include partial connectivity of networks as can be seen in many types
of ad-hoc wireless networks with frequent network partitions, long
propagation time as experienced in inter-planetary and deep space
networks, and frequent link disruptions due to the mobility of nodes
as observed in terrestrial wireless network
environments. Experimenting network architectures, protocols, and
mobility models in such real-world scenarios is difficult due to the
complexities involved in the network environment. Therefore, in this
document, we present the documentation of an Urban Delay Tolerant
Network Simulator (UDTNSim) version 0.1, capable of simulating urban
road network environments with DTN characteristics including mobility
models and routing protocols. The mobility models included in this
version of UDTNSim are ({\it i})~Stationary Movement, ({\it
  ii})~Simple Random Movement, ({\it iii})~Path Type Based Movememt,
(iv) Path Memory Based Movement, ({\it v})~Path Type with Restricted
Movement, and ({\it vi})~Path Type with Wait Movement. In addition to
mobility models, we also provide three routing and data hand-off
protocols: ({\it i})~Epidemic Routing, ({\it ii})~Superior Only
Handoff, and ({\it iii})~Superior Peer Handoff. UDTNSim~v0.1 is
designed using object-oriented programming approach in order to
provide flexibility in addition of new features to the DTN
environment. UDTNSim~v0.1 is distributed as an open source simulator
for the use of the research community.


\tableofcontents
\phantomsection
\addcontentsline{toc}{chapter}{List of Figures}
\cleardoublepage
\listoffigures

\cleardoublepage
\setcounter{page}{0}
\pagenumbering{arabic}
\chapter{Introduction}\label{chp:intro}
Delay Tolerant Networking~(DTN)\index{Delay Tolerant
  Networking}\index{DTN} is the area of networking which can address
the challenges in disconnected and disrupted networks without an
end-to-end connection~\cite{fall2003delay}. Real world wireless
networks experience delays due to several factors such as long
transmission time, obstacles in line-of-sight, and disruption of the
links resulting from low transmission range and node mobility. Delay
due to long transmission time occurs in interplanetary communications,
whereas, frequent link-breaks (disruptions) are the major cause of
end-to-end delay in mobile ad-hoc networks such as vehicular networks.

An architecture for DTN~\cite{fall2003delay} is proposed with an
addition of a \textit{bundle layer}\index{Bundle layer} between the
application and transport layer of traditional OSI model. A protocol
known as \textit{bundle protocol}~\cite{RFC5050} is designed to manage
the operations of bundle layer. The layer creates its own data units,
known as \textit{bundles}\index{Bundles}, from the data it received
from the application layer. In case of significant delays, the bundle
layer stores the bundles until it gets an opportunity to handover the
data to the destination or to a node with a higher chance of
delivering the message toward the destination. That is, the bundle
layer uses a store-and-forward approach, known as
\emph{custody-transfer}\index{Custody-transfer}, in transferring the
data towards the subsequent nodes. In order to achieve reliability in
communication, the node to which the data is transferred takes the
responsibility of \emph{custodian} of the message so that the message
can be delivered without loss.

Several routing protocols are developed for efficient communication in
DTNs. Some of them include epidemic
protocol~\cite{vahdat2000epidemic}, spray and
wait~\cite{spyropoulos2005spray}, MaxProp~\cite{burgess2006maxprop},
and PROPHET~\cite{lindgren2003probabilistic}. The protocols use
different metrics such as delivery probability of nodes, history of
nodes' contacts, the time for delivery, and buffer size for computing
the routing decision. In general, the protocols are designed to
maximize the delivery probability while minimizing the transmission
time, data loss, and communication overhead.

Similar to routing, in case of networks in terrestrial environments
such as vehicular networks, mobility pattern of the nodes has
significant influence on the network delay. The mobility pattern
varies with different networks. Examples of different mobility
patterns include mobility of vehicles in a road network, mobility of
boats in the sea, and mobility of drones. The quality of routing
decisions depends on how accurately we can predict the mobility
pattern of nodes and, thereby, the contacts between them. Random
mobility makes the task more challenging. Moreover, there occur
situations where we need to design mobility pattern if we have
sufficient information about the terrain. For example, in a disaster
environment, we need to design efficient mobility patterns for the
emergency response personnel to respond to emergency calls with
available resources. Therefore, mobility models are very important
aspect in delay tolerant networks.

Experimenting mobility models and DTN protocols in a real world set-up
may not be feasible always. Simulations provide better options for
testing and evaluating new mobility models and routing
protocols. Here, we discuss an Urban Delay Tolerant Network
Simulator~(UDTNSim)~\cite{udtnsim}\index{UDTNSim}\footnote{The source
  code of UDTNSim is available at
  \href{https://github.com/iist-sysnet/UDTNSim}{\tt
    https://github.com/iist-sysnet/UDTNSim}}, developed using
Python-2.7~\cite{python27}, which is capable of creating DTN
environments with real world node mobility and associated routing
protocols. The simulator provides considerable flexibility in adding
modules to customize as per our requirements.

\chapter{UDTNSim Architecture}\label{chp:approach}
\begin{figure}[h]
  \centering
  \includegraphics[scale=0.55]{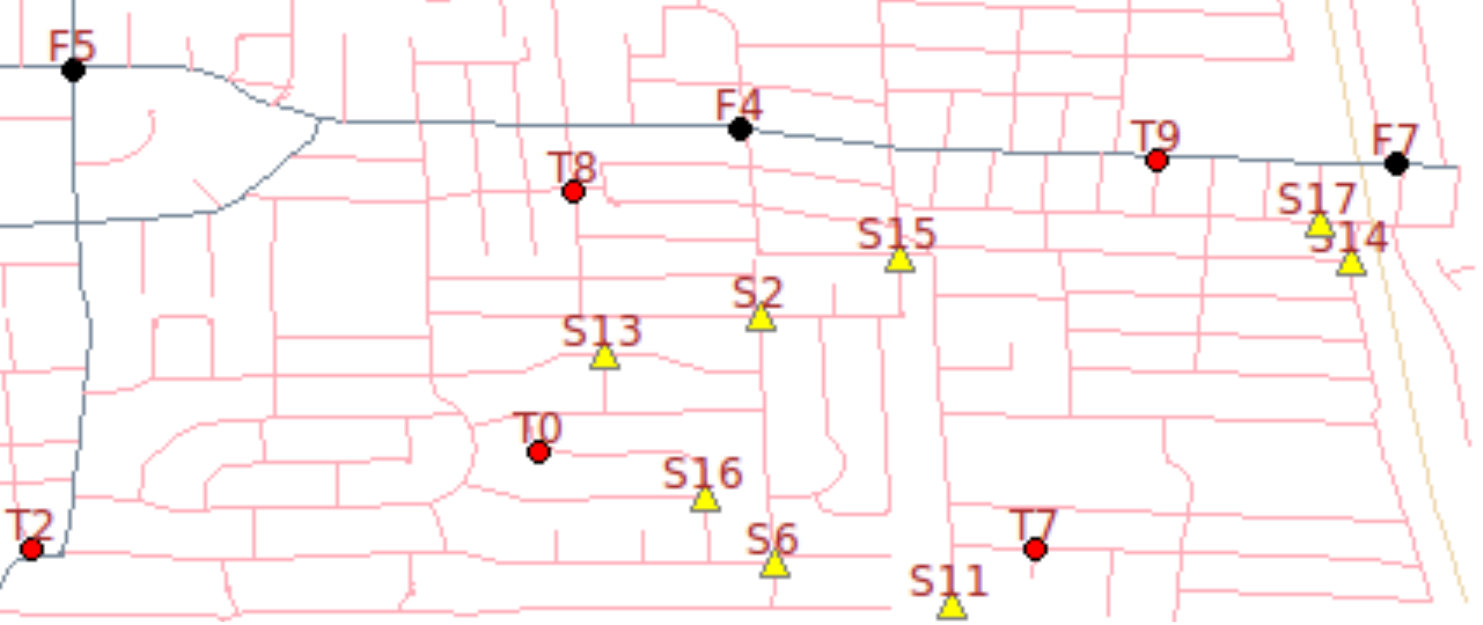}
  \caption{A sample simulation scenario.}
  \label{fig:scenario}
\end{figure}

The simulation environment is defined using a geographical area with a
road network and objects~(stationary well as mobile). The stationary
objects include sensors, wireless routers, and other devices which
provide wireless connectivity to objects. Mobile agents (for example
bikes, cars, humans, and drones) connected with wireless devices are
represented as mobile objects. Stationary objects can be positioned
(either at random or using custom algorithms) in the specified
geographical area and mobile nodes move through the area either
through roads, in case of vehicles, or through air, in case of drones,
and communicate with other mobile nodes as well as stationary
nodes. The purpose of network includes data gathering from sensors and
communication between different kinds of
objects. Figure~\ref{fig:scenario} shows an example scenario with a
road network and set of stationary and mobile objects.

\begin{figure}[h]
  \centering
  \includegraphics[scale=0.35]{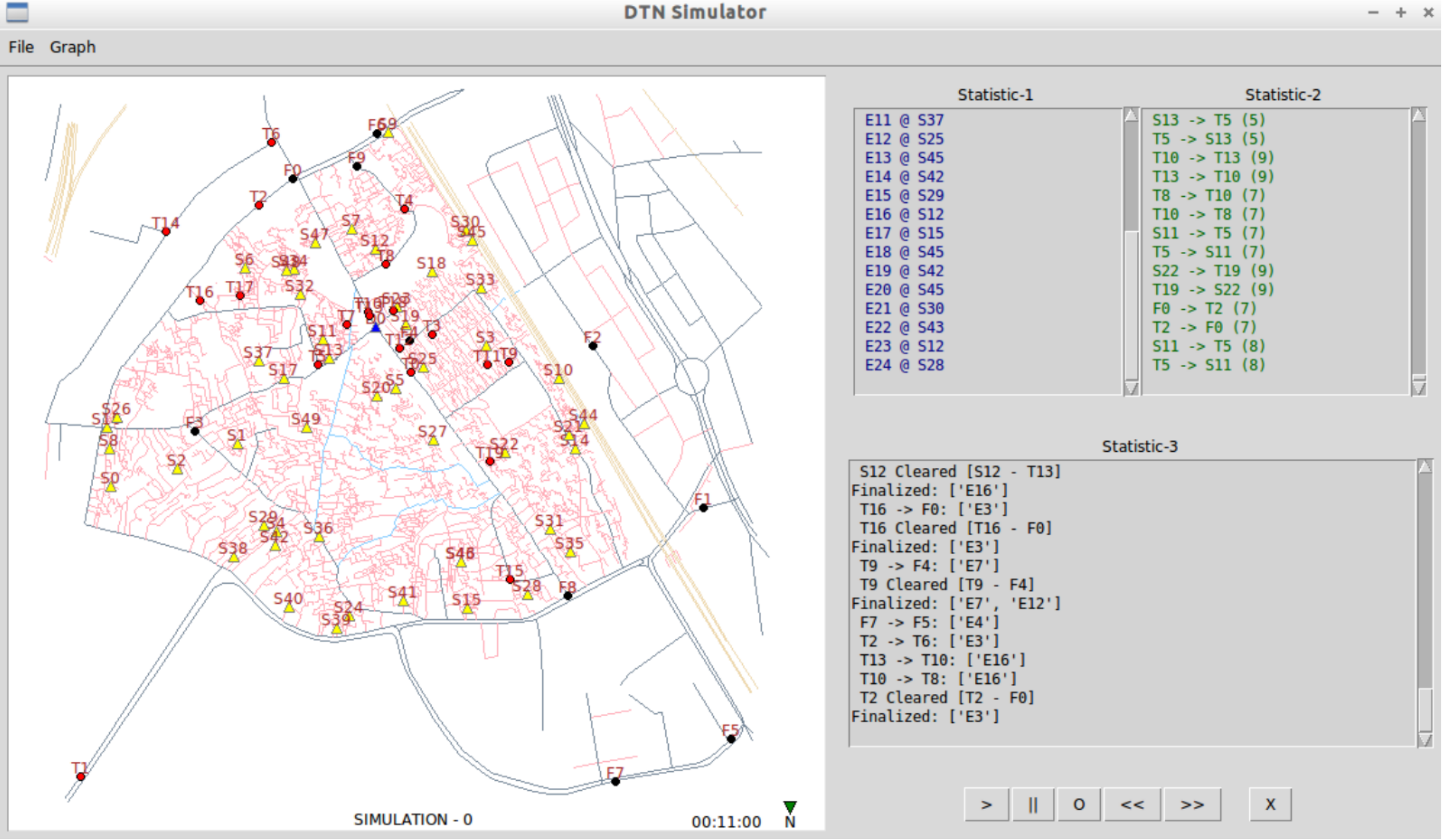}
  \caption{A snapshot of UDTNSim using the map of Dharavi, India.}
  \label{fig:scenario1}
\end{figure}

In Figure~\ref{fig:scenario}, highways\index{Highways} are represented
as black lines and the roads which are accessible for two-wheeler
objects as red lines. Yellow triangles represent stationary nodes
(sensors) which are distributed in the geographic area. The mobile
objects are represented using filled circles~(black circle represents
four-wheeler vehicles and red, the two-wheeler vehicles). From
Figure~\ref{fig:scenario}, it is clear the four-wheeler vehicles can
only access the highways~(black lines) and two-wheeler vehicles can
access both highways as well as two-wheeler ways~(red lines). The
mobile nodes can move through the road network and collect data from
the stationary nodes. A moving object can communicate with another
when they come in the range of each other, i.e., while they are in
\textit{contact}\index{Contact}. If the destination node for a message
is not in the transmission range, the source can forward the message
to an object which has a higher chance of meeting with the
destination. A simulation snapshot of UDTNSim using the map of shanty
town Dharavi, India, is shown in Figure~\ref{fig:scenario1}. Since the
simulation is based on the real world scenario, we use object oriented
programming approach for its implementation.

\section{A Modular View of UDTNSim}\label{sec:simarch}
\begin{figure}[t!]
  \centering
  \includegraphics[width=0.75\textwidth]{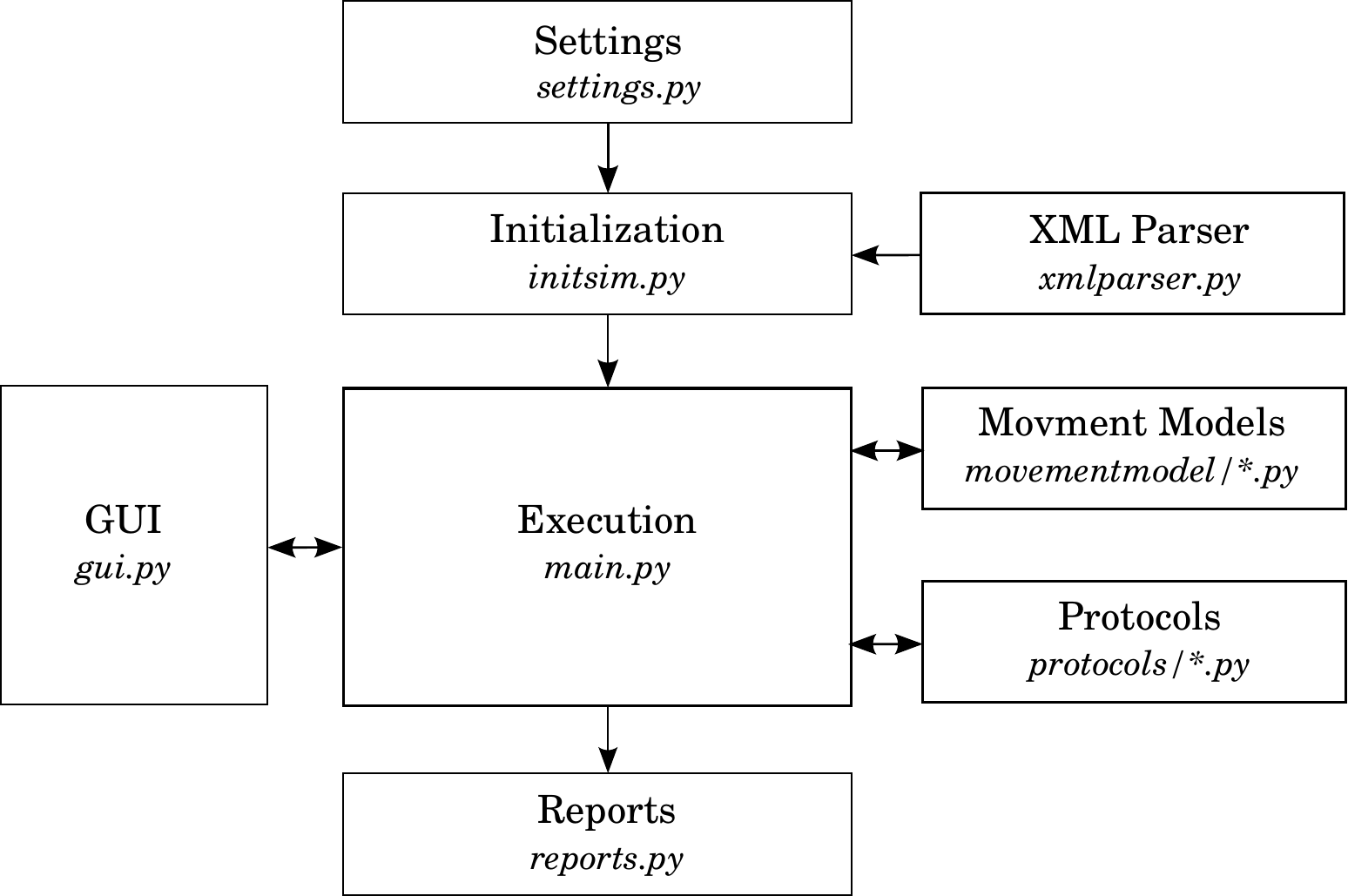}
  \caption{UDTN simulator architecture.}
  \label{fig:simarch}
\end{figure}

The UDTN simulator is divided into eight modules to simplify the
process of adding new features or modifying the existing
ones. Figure~\ref{fig:simarch} shows the architecture of the simulator
with different modules and their interactions.
\begin{enumerate}
\item \texttt{Settings:}\index{Settings} Module to provide the
  simulation environment. It includes a configuration file
  `\texttt{sim.config}\index{Files!sim.config@\texttt{sim.config}}' as
  well as the libraries to parse the configuration file.
\item \texttt{XML Parser:}\index{XML Parser} Module to parse the
  geographical location from the map information (which is in XML
  format extracted from \texttt{OpenStreetMap}\cite{OpenStreetMap})
  and convert it into simulation variables.

\item \texttt{Initialization:}\index{Initialization} The module
  initializes the simulator environment using the configuration
  provided by the user as well as the map information.
\item \texttt{Movement Models:}\index{Mobility Model} Handles
  different movement models for the objects.

\item \texttt{Protocols:}\index{Protocols} Defines routing protocols
  for the objects.
\item \texttt{Execution:}\index{Execution} Coordinates different modules and execute
  the simulation. Execution module communicates with \texttt{Movement
    Models, Protocols, GUI,} and \texttt{Reports} modules.
\item \texttt{GUI:}\index{GUI} Handles the graphical user interface of
  the simulator.
\item \texttt{Reports:}\index{Reports} Creates of simulation-related
  logs.
\end{enumerate}

\section{Implementation of Modules}\label{sec:oops}
Each component of the simulator is implemented using objected oriented
programming\index{Object Oriented Programming} approach to make the
code reusable and provide flexibility. For example, the basis of
movement model module is a class to represent stationary objects with
basic attributes such as a unique ID, its geographical position, and a
buffer to store data. The concept of mobility is introduced from
stationary object with the addition of functions required for the
position change, i.e., the mobility. In other words, the class for a
mobile object can be defined by inheriting properties from the
stationary object. In similar way, different kinds of mobility can be
defined from the mobility class with redefinition of the function for
position change. Therefore, object oriented programming approach
provide the developers more flexibility in adding new features with
minimum effort.

\section{Simulator Configuration}\label{sec:settings}
For any simulator, the execution environment is specified in terms of
simulation parameters. As we discussed at the beginning of this
chapter, the simulator consists of a geographical area and
objects~(both stationary and mobile). We need to specify the
parameters required for each of these components. For example, the
geographical area is specified as a map~(represented using XML). The
map can be exported from
\texttt{www.openstreetmap.org}\cite{OpenStreetMap}, in \texttt{.osm}
format. The parameters for a mobile object include ID, the road-types
it can travel, buffer size, mobility model, and routing protocol.

In general, the parameters are specified using a configuration
file. We use a file
\texttt{sim.config}\index{Files!sim.config@\texttt{sim.config}} to
specify the environmental parameters (such as simulation name,
simulation time, map information, and report directory) and the
movement object specific parameters~(such as group label, transmission
range, buffer size, and speed).

\subsection{Simulation Parameters}\label{subsec:simparams}Simulation
parameters are classified into two: (\textit{i})~general parameters and
(\textit{ii})~node specific parameters.
\index{Parameters}
\subsubsection{General Parameters}\label{subsubsec:genparams}General
parameters include those which are required for every component in the
simulator.

\begin{enumerate}
\item \texttt{Simulation\_Name}: Provides a name for the simulation.

\item \texttt{No\_of\_Simulations}:
  \index{Parameters!No\_of\_Simulations@\texttt{No\_of\_Simulations}}In
  most cases, especially in cases where randomness is involved, we
  need to run the simulations multiple times and aggregate the
  observations obtained from each of them. The parameter
  \texttt{No\_of\_Simulations} allows to set the required number of
  simulations.

\item
  \texttt{Simulation\_Time}:\index{Parameters!Simulation\_Time@\texttt{Simulation\_Time}}
  Represents the \textit{real world} simulation time specified in
  hours.

\item \texttt{Map}:\index{Parameters!Map@\texttt{Map}} Represents the
  map of geographical area in \texttt{.osm} format. The file should be
  specified using its path, either absolute or relative. By default,
  the simulator has a directory \texttt{maps}, where \texttt{.osm}
  files for different geographical areas can be placed.

\item
  \texttt{Report\_Directory}:\index{Parameters!Report\_Directory@\texttt{Report\_Directory}}
  Represents the location for placing the simulation logs. The
  directory can be specified using an absolute or relative path. At
  present, the simulator has a directory \texttt{logs} for
  simulation-related logs. In the current version, separate
  sub-directories are created for each of the objects (See
  Section~\ref{sec:reports}).

\item
  \texttt{GUI\_Enabled}\index{Parameters!GUI\_Enabled@\texttt{GUI\_Enabled}}:
  Enables the Graphical User Interface~(GUI). It has two values,
  \texttt{True} or \texttt{False}. GUI can be enabled for testing the
  movement models and protocols. Since GUI consumes more CPU time for
  simulation, it is recommended to make \texttt{GUI\_Enabled~=~False}
  for simulations which take significant time.

\item
  \texttt{Path\_Types}\index{Parameters!Path\_Types@\texttt{Path\_Types}}:
  Since the simulation environment has a geographical area, we need to
  specify the types of roads that we consider for our simulation. The
  mobile objects move through the specified types of roads and
  communicate with other objects. We need to assign unique numbers for
  each of the road-types so that the numbers can be used to specify
  the allowed road-types for each object.
  
\item
  \texttt{Random\_Msg\_Gen\_Parameter}\index{Parameters!Random\_Msg\_Gen\_Parameter@\texttt{Random\_Msg\_Gen\_Parameter}}:
  Represents the event generation parameter. Its value is a tuple of
  the form $[m, n]$, i.e., $m$ events are generated in every $n$
  hours.

\item
  \texttt{No\_of\_Hosts\_Groups}:\index{Parameters!No\_of\_Hosts\_Groups@\texttt{No\_of\_Hosts\_Groups}}
  Indicates the number of object groups present in the simulation. If
  we have more than one object with same specifications, we can
  organize them as an host-group. Further, we need to describe the
  specifications for each group.

\end{enumerate}

\subsubsection{Group Specific Parameters}\label{subsubsec:grpparams}Here,
we describe the parameters specific for each object group. Every
object in a group are assigned with the same parameters specified for
that group.
\begin{enumerate}
\item
  \texttt{Group\_ID}:\index{Parameters!Group\_ID@\texttt{Group\_ID}}Unique
  ID for the group.

\item \texttt{Label}:\index{Parameters!Label@\texttt{Label}} Label for
  distinguishing each group in the simulation GUI. If \texttt{Label =
    T}, the objects in the group are assigned with names \texttt{T1,
    T2, T3}, and so on.

\item \texttt{Paths}\index{Parameters!Paths@\texttt{Paths}}: Assign the road-types
  through which the objects in a group can travel. We need to give the
  list of numbers corresponding to the path-types given in the
  parameter \texttt{Path\_Types}, specified in
  Section~\ref{subsubsec:genparams}.

\item
  \texttt{No\_of\_Hosts}:\index{Parameters!No\_of\_Hosts@\texttt{No\_of\_Hosts}}
  The number of nodes present in the group.

\item
  \texttt{TX\_Range}:\index{Parameters!TX\_Range@\texttt{TX\_Range}}
  Transmission range of the node specified in meters.

\item
  \texttt{Buffer\_Size}:\index{Parameters!Buffer\_Size@\texttt{Buffer\_Size}}
  Size of the buffer in each node specified in Kilobytes~(K),
  Megabytes~(M), or Gigabytes~(G). At present the buffer size is of
  infinite size.

\item \texttt{Speed}:\index{Parameters!Speed@\texttt{Speed}}Speed of
  the node in kilometer per hour (km/hr).

\item \texttt{Mobile}:\index{Parameters!Mobile@\texttt{Mobile}}
  Parameter that distinguishes a mobile node from a stationary
  node. It takes the value either \texttt{True} or
  \texttt{False}. \texttt{True} represents a mobile node while
  \texttt{False}, a stationary node.

\item \texttt{Movement}:\index{Parameters!Movement@\texttt{Movement}}
  Represents the mobility model for the nodes. Different movement
  models such as \texttt{StationaryMovement, SimpleRandomMovement,}
  and \texttt{PathTypeMovement} are defined in the directory
  \texttt{movementmodel}.

\item
  \texttt{Junction\_Delay}:\index{Parameters!Junction\_Delay@\texttt{Junction\_Delay}}
  Delay that occurred at the road junctions specified in seconds.

\item \texttt{Color}: Color of the nodes in GUI.

\item \texttt{Protocol}\index{Parameters!Protocol@\texttt{Protocol}}:
  Specifies the routing protocol for the nodes. For example, a node
  following \texttt{EpidemicHandoff} transfers its data to the nodes
  which comes in contact with it. In case of
  \texttt{SuperiorOnlyHandoff}, a node transfers its data only to a
  node which is higher at the classification hierarchy. A node which
  acts as Depot do not transfer the data to other nodes, however, it
  only receives the data from other nodes. The protocols are defined
  in the subdirectory \texttt{protocols}.
\end{enumerate}
\index{Variables}
\index{Files}

\subsection{Parser}\label{subsec:setparser}The code for parsing the
configuration file \texttt{sim.config} is written in
\texttt{parser/settings.py}\index{Files!settings.py@\texttt{settings.py}}. Since
the first step of simulation is to read the environment settings, an
object of type \texttt{Settings} is created at the start of simulation
(in \texttt{initsim/init.py}).
\begin{lstlisting}[style=customstyle]
  envt_params = settings.Settings('sim.config')
\end{lstlisting}
The class \texttt{Settings}\index{Class!Settings@\texttt{Settings}}
contains two dictionaries,
\texttt{envt}\index{Variables!envt@\texttt{envt}} and
\texttt{groups}\index{Variables!groups@\texttt{groups}}, which
represent general parameters and group specific parameters,
respectively. The dictionary \texttt{envt} has the parameter name as
the key and the value as the value given by the user. For
\texttt{groups}, key is the \texttt{Group\_ID} and value as another
dictionary with key as parameter name and value the corresponding node
specific value. The function \texttt{read\_settings()} reads the
entire content from the settings file,
\texttt{sim.config}\index{Files!sim.config@\texttt{sim.config}}, and
populate the dictionaries \texttt{envt} as well as
\texttt{groups}. Logical view of a \texttt{Settings} object is shown
in Figure~\ref{fig:settings}.

\begin{figure}[ht!]
  \centering \texttt{
    \begin{tabular}{| c | p{8cm} |}
      \hline
      \textbf{Attribute} & \textbf{Value} \\\hline
      settings\_file & `sim.config' \\ \hline
      envt & \{`Simulation\_Name':`SampleDTN', `GUI\_Enabled':False,
             \ldots \}\\ \hline 
      groups & \{`G1':\{`Label':`T', `Mobile':True, \ldots \}, `G2':\{\ldots\}, \ldots \}\\\hline
    \end{tabular}}
  \caption{Example view of the \texttt{settings} object}
  \label{fig:settings}
\end{figure}

\subsection{Addition of New Simulation
  Parameters}\label{subsec:addsimparams}
As the simulation needs change, there may be a requirement for new set
of parameters for the environment as well as for the nodes. Therefore,
the simulator provides a provision to add new parameters using two
files \texttt{envt\_params.in} and \texttt{group\_params.in} in the
subdirectory \texttt{parser}\footnote{The source code for UDTNSim
  Parser module is available at\\
  \href{https://github.com/iist-sysnet/UDTNSim/tree/master/parser}{\tt
    https://github.com/iist-sysnet/UDTNSim/tree/master/parser}}. For
the addition of a new parameter, we need to append a line of the form
\texttt{ParameterName:Type} to the file \texttt{envt\_params.in} for a
general parameter, or to \texttt{group\_params.in} for a group
specific parameter. If the value corresponding to a parameter is a
list with items which are of non-string type, we may need to convert
it into appropriate type in the code (either in \texttt{settings.py}
or wherever we access that parameter).  \index{Functions}
\subsection{XML Parser}\label{subsec:xmlparser}
We consider the map input as an XML file, in \texttt{.osm} format,
exported from \texttt{openstreetmap.org}~\cite{OpenStreetMap}. In
order to extract the information required for the simulation, we
define three functions in \texttt{parser/xmlparser.py} file.
\begin{enumerate}
\item
  \texttt{parse\_osm():}\index{Functions!parse\_osm@\texttt{parse\_osm}}
  This function extracts the map information given in \texttt{.osm}
  file into two dictionaries, \texttt{node\_dict} and
  \texttt{way\_dict}. \texttt{node\_dict} keeps the information of
  geographical points with their latitude, longitude, corresponding
  cartesian coordinates for GUI, and the identifiers of the roads in
  which the geographical point takes part in.
  \begin{lstlisting}[style=customstyle]
    node_dict = {node_id: [(lat, lon), (y, x), [w1, w2, ...], ...}
  \end{lstlisting}
  Similar to \texttt{node\_dict}, \texttt{way\_dict} stores the
  information related to ways~(roads). The key is the road identifier
  and value is a list with node IDs~(keys from \texttt{node\_dict}) of
  the geographical points which create the way, the type of the road,
  and the length of the road~(in kilometers).
  \begin{lstlisting}[style=customstyle]
    way_dict = {way_id: [[node_id1, node_id2,...], way_type, way_length], ...}
  \end{lstlisting}
  The function takes inputs the \texttt{osm} file and the
  dimension~(height and width) of the simulator GUI. The function is
  called from \texttt{init\_sim\_envt()} in
  \texttt{init/initsim.py}. The map file is taken from the path
  provided in \texttt{sim.config} using \texttt{Map} parameter.
  \begin{lstlisting}[style=customstyle]
    Map = maps/manhattan.osm
  \end{lstlisting}

\item
  \texttt{traverse\_way():}\index{Functions!traverse\_way@\texttt{traverse\_way}}
  The XML file extracted from from \texttt{OpenStreetMap} consists of
  two tags for representing roads ({\it i})~\texttt{<node>} and ({\it
    ii})~\texttt{<way>}. The node information (i.e.,the geographical
  points) from \texttt{<node>} tag is read by the function
  \texttt{parse\_osm()}. We define the function
  \texttt{traverse\_way()} to extract way information given in
  \texttt{<way>} tags. The function is invoked from the
  \texttt{parse\_osm()} function when it interprets a line with a
  \texttt{<way>} tag.
  \begin{lstlisting}[style=customstyle]
    # Detected a `way' tag. The function `traverse_way' is 
    # called to retrieve  the way information 
    elif line.find(``<way'') + 1: 
         components = line.split()
         way_id = int(components[1].split(``\"")[1]) 
         node_order, way_type, file_ptr = traverse_way(file_ptr, node_dict, way_id) 
         # Add way information to way_dict 
         if way_type: 
             way_dict[way_id] = [node_order, way_type]
  \end{lstlisting}
\item
  \texttt{normalize\_map():}\index{Functions!normalize\_map@\texttt{normalize\_map}}
  The XML file from \texttt{OpenStreetMap} provides the way
  information, including the geographical points that make the
  way~(roads). The information of road intersections, which is a
  pre-requisite for representing road network as a graph, is not
  provided with the XML file. Therefore, the function
  \texttt{normalize\_map()} takes \texttt{node\_dict} and
  \texttt{way\_dict} as inputs and finds the road
  intersections. Further, the ways are divided according to the number
  of intersections and the dictionaries are modified accordingly.
\end{enumerate}

In order to add a parser for a new type of map file, appropriate functions
should be defined in the directory \texttt{parser}. The functions should parse
the file and generate two dictionaries, \texttt{node\_dict} and
\texttt{way\_dict}, with the same structure as discussed in the beginning of
this section.  

\section{Summary}\label{sec:chp2summary}
In this chapter, we discussed the architecture of the UDTN simulator,
its configuration, and the parser modules for input parameters. The
implementation details of the remaining modules are described in the
following chapters.

\chapter{Movement Models}\label{chp:movementmodels}
Movement models\index{Movement model} define the mobility pattern of
the objects. A classical example is random way-point
model~\cite{bai2004survey}\index{Random way-point model} where, the
node randomly chooses a destination and moves toward it. Once reaching
the destination, the node again selects a random destination and the
process continues.

Since UDTN simulator considers movement of nodes through a road
network, we need to consider the characteristics of each road as well
as each object while defining the mobility. For instance, road-type
and the type of object~(vehicle) are important factors while defining
the mobility. That is, four-wheeler nodes such as cars and buses can
only travel through the road with sufficient width. For example, cars
can easily traverse through a highway, not through a two-wheeler road
or a footpath. However, a pedestrian can travel through all types of
roads.

In our simulator, we define three movement models: ({\it
  i})~\emph{Stationary Movement}, ({\it ii})~\emph{Simple Random
  Movement}, ({\it iii})~\emph{Path Type Based Movememt}, ({\it
  iv})~\emph{Path Memory Based Movement}, ({\it v})~\emph{Path Type
  with Restricted Movement}, and ({\it vi})~\emph{Path Type with Wait
  Movement}\footnote{The source code for different movement models
  implemented in UDTNSim is available at\\
  \href{https://github.com/iist-sysnet/UDTNSim/tree/master/movementmodel}{\tt
    https://github.com/iist-sysnet/UDTNSim/tree/master/movementmodel}}.

\section{Stationary Movement}\label{sec:stationarymovement}
As the name implies, stationary movement\index{Stationary movement} is
defined as \emph{no} movement. Stationary movement is defined
especially for static nodes such as sensors and fixed wireless routers
or access points. The movement model is defined as a class
\texttt{Stationary}\index{Class!Stationary@\texttt{Stationary}} in the
file
\texttt{stationary.py}\index{Files!stationary.py@\texttt{stationary.py}}. Also,
the purpose of this movement model is to define the characteristics of
an object such as identity, group identity, position~(geographic
position and its corresponding pixel position in the GUI), storage
buffer, and a report mechanism for creating logs. The attributes
defined for stationary object include the following:

\begin{enumerate}
\item \texttt{obj\_id:} Unique identifier for the object.
\item \texttt{group\_id:} Identifier of the group where the object
  belongs. More than one movement objects can have the same group
  identifier.
\item \texttt{prev\_node:} Current vertex\footnote{The terms
    \textit{vertex} and \textit{junction} are used interchangeably in
    the document.} identifier~(a key from \texttt{node\_dict}) of the
  node, i.e., the junction the object recently visited.
\item \texttt{next\_node:} The vertex toward which the object is
  moving. In stationary movement, \texttt{next\_node} is the same as
  the \texttt{prev\_node}.
\item
  \texttt{curr\_geo\_pos:}\index{Variables!curr\_geo\_pos@\texttt{curr\_geo\_pos}}
  The current geographical position of the node. It is a tuple of the
  form \texttt{(lat, lon)} where \texttt{lat} and \texttt{lon}
  represent latitude and longitude, respectively.
\item \texttt{curr\_pix\_pos:} The current pixel position of the
  object in the GUI. It is also a tuple of the form \texttt{(y, x)}.
\item \texttt{buffer:} A list for storing data.
\item
  \texttt{report\_obj:}\index{Variables!report\_obj@\texttt{report\_obj}}
  An object of
  \texttt{MovementReport}\index{Class!MovementReport@\texttt{MovementReport}}
  class to create logs related to the object.
\item
  \texttt{protocol\_obj:}\index{Variables!protocol\_obj@\texttt{protocol\_obj}}
  An object of the routing protocol. Different routing protocols are
  discussed in Chapter~\ref{chp:protocols}.
\end{enumerate}

The class
\texttt{Stationary}\index{Class!Stationary@\texttt{Stationary}} has a
member function \texttt{compute\_initial\_node()}, which randomly
chooses a vertex in the geographical location to position the
stationary object. That is, the vertex chosen is assigned to the
\texttt{prev\_node} attribute of stationary node. The function
\texttt{compute\_initial\_node()}\index{Functions!compute\_initial\_node@\texttt{compute\_initial\_node}}
can be overridden to position nodes at specific positions as required
for the simulation.

\section{Simple Random Movement}\label{sec:simplerandommovememt}
Simple random movement is an adoption of random way-point model on to
a road network. Here, unlike stationary movement, the object moves
from one vertex~(road junction) to the adjacent vertex through the
road~(edge) connecting them. We define simple random movement model as
follows: a node at a vertex randomly chooses a vertex from its
adjacent vertices~(excluding the adjacent vertex that it comes from)
and moves toward the selected vertex through the edge connecting those
two vertices. If the vertex is an end point of a road, the object
returns back to its previous vertex.

Simple random movement is defined as a class
\texttt{SimpleRandomMovement}\index{Class!SimpleRandomMovement@\texttt{SimpleRandomMovement}}
in file
\texttt{simplerandom. py}\index{Files!simplerandom.py@\texttt{simplerandom.py}}. The
movement model inherits all the properties of stationary node from
\texttt{Stationary}\index{Class!Stationary@\texttt{Stationary}}
class. Further, we define the following additional attributes to
realize the movement:
\begin{enumerate}
\item \texttt{curr\_way:} The way id~(key from the dictionary
  \texttt{way\_dict}) of a way where the object locates.
\item \texttt{speed:} Speed of the object in km/hr.
\item
  \texttt{ways\_visited:}\index{Variables!ways\_visited@\texttt{ways\_visited}}
  A list to store the \texttt{way\_id}s of the ways the object already
  visited.
\item \texttt{time\_traveled:} The travel time of the object in
  seconds.
\item \texttt{mvmt\_points:} A list to keep the movement points
  (geographical as well as pixel points) of the way on which the node
  is moving. For example, if a node moves from vertex
  A~(\texttt{prev\_node}) to vertex B~(\texttt{next\_node}), the
  geographical points between~A and~B are computed and stored in
  \texttt{mvmt\_points}. At each time step, the node moves from one
  point to the next.
\item \texttt{mvmt\_pt\_index:} An index to the list
  \texttt{mvmt\_points}, which represents the current position of the
  node.
\end{enumerate}

Apart from the movement specific attributes, we define three functions
to handle the movement:

\begin{enumerate}
\item
  \texttt{update\_position():}\index{Functions!update\_position@\texttt{update\_position}}
  At each time step, the function \texttt{update\_position()}, changes
  the position of object through the points specified in
  \texttt{mvmt\_points}~(geographical positions of a road). If the
  object reaches the last position in \texttt{mvmt\_points}, i.e., end
  of the current way, the object needs to choose an adjacent vertex at
  random, which is done using the function
  \texttt{compute\_next\_node()}. If the current position is not the
  end of the \texttt{curr\_way}, the node is shifted to the next
  position in \texttt{mvmt\_points} using an increment to
  \texttt{mvmt\_pt\_index}.
\item
  \texttt{compute\_next\_node():}\index{Functions!compute\_next\_node@\texttt{compute\_next\_node}}
  The function chooses an adjacent vertex~(junction) from neighboring
  vertices at random. If the vertex is an end vertex, then the
  function considers the previous vertex as the next vertex. The next
  way~(link) for the node will be the way connecting the current
  vertex and next chosen vertex.
\item
  \texttt{populate\_way\_points():}\index{Functions!populate\_way\_points@\texttt{populate\_way\_points}}
  After choosing the next link to travel, we need to find the
  geographical points between the \texttt{prev\_node} and
  \texttt{next\_node}. In order to trace the points that make the
  link, we interpolate the points using the geographical positions
  available~(from \texttt{.osm} file) in the \texttt{curr\_way}. The
  available geographical points can be assessed using the dictionary
  \texttt{way\_dict[curr\_way][0]}. The interpolated points are stored
  in \texttt{mvmt\_points} and the index variable
  \texttt{mvmt\_pt\_index} is reset to 0. If any junction delay is
  specified for the object, we append required number of points~(same
  as the last point in \texttt{mvmt\_points}) at the end of
  \texttt{mvmt\_points}.
\end{enumerate}

\texttt{SimpleRandomMovement}\index{Class!SimpleRandomMovement@\texttt{SimpleRandomMovement}}
is the simplest mobility model for the nodes. The model is more
suitable if the road network has only a single type of road or the
mobile objects are capable of moving through every type of
roads. Here, the movement objects do not consider the type of road for
their movement. In order to simulate a more realistic model, it is
necessary to consider the types of roads as well as the types of
movement objects. \texttt{Path Type Based Movement} is suitable
direction toward it.

\section{Path Type Based Movement Model}\label{sec:pathtypemovement}
\index{Path Type Based Movement} Simple random movement may not be
realistic if we consider a road network with different types of roads
and different types of mobile objects. That is, simple random movement
does not distinguish between different types of roads as well as
different types of mobile nodes. As explained in the example at the
beginning of this chapter, we need to consider the type of vehicle and
type of roads it can traverse. Therefore, we define a path-type based
movement model where each mobile object chooses an adjacent neighbor
vertex only if it has a road with sufficient width to travel toward
the destination. For example, a four-wheeler chooses a neighboring
vertex only if there exists a highway connecting to that
vertex. Therefore, path-type based movement model can be considered as
simple random movement with constraints on road-type.

The movement model is defined as a class
\texttt{PathTypeMovement}\index{Class!PathTypeMovement@\texttt{PathTypeMovement}}
in the file \texttt{path
  type.py}\index{Files!pathtype.py@\texttt{pathtype.py}}. Since, the
mobile node shares all the properties of simple random
movement~(except the vertex selection procedure), the class
\texttt{PathTypeMovement} is defined as a class derived from
\texttt{Simple
  RandomMovement}\index{Class!SimpleRandomMovement@\texttt{SimpleRandomMovement}}. Here,
we need to redefine the member functions \texttt{compute\_initial
  \_node()} and \texttt{compute\_next\_node()} to override the
functions in simple random movement.
\begin{enumerate}
\item
  \texttt{compute\_initial\_node():}\index{Functions!compute\_initial\_node@\texttt{compute\_initial\_node}}
  When a mobile object gets created, it is positioned at a random
  vertex in the graph. In path-type movement model, we need to
  randomly select a vertex with roads attached to it having the type
  that can be traversed by that object. That is, for computing an
  initial vertex for a four-wheeler, we need to find a vertex with at
  least one highway attached to it. In this function, first we find
  the possible road-types (given in the
  \texttt{sim.config}\index{Files!sim.config@\texttt{sim.config}}
  file) that the node can traverse using the following statement.
  \begin{lstlisting}[style=customstyle]
    possible_paths = global_params.envt_params.groups\
    [self.group_id]['Paths']
  \end{lstlisting}
  Further we choose a vertex at random from the graph and checks any
  of its neighbor is connected with a road-type given in
  \texttt{possible\_paths}. If none of the edges are with required
  type, then the process is repeated with another randomly chosen
  vertex.
\item
  \texttt{compute\_next\_node():}\index{Functions!compute\_next\_node@\texttt{compute\_next\_node}}
  Similar to simple random movement model, when a node reaches a
  vertex/junction, it collects the neighbor vertices. Among them, the
  node creates a list of neighbor vertices~(\texttt{possible\_nodes})
  which are connected by any of the road-types given in its
  \texttt{possible\_types}. From \texttt{possible\_nodes}, a vertex is
  chosen at random as the \texttt{next\_node}. If the list
  \texttt{possible\_nodes} is empty, \texttt{next\_node} is assigned
  with the \texttt{prev\_node}.
\end{enumerate}

\section{Path Memory Based Movement Model}
\label{sec:pathmemeory}

Path Memory Based Movement Model\index{Path Memory Based Movement} is
a path-type based model which uses the history of nodes' traversal for
taking decision on further movement. The purpose of path-memory model
is to avoid the selection of already traversed roads in the map and,
thereby, explore maximum possible locations in the geographical
area. Here, the function \texttt{compute\_next\_node()} is redefined
to make the movement memory-based. After computing
\texttt{possible\_nodes} as in path-type movement model, the function
discards already traversed ways using the list
\texttt{ways\_traversed}.
\begin{lstlisting}[style=customstyle]
  untraversed_nodes = [] for node in possible_nodes: if
  road_graph[self.next_node][node]['way_id'] not in self.ways_visited:
  untraversed_nodes.append(node)
                
  if untraversed_nodes: possible_nodes = untraversed_nodes[:]
\end{lstlisting}
In case of no untraversed roads connected to a junction, the algorithm
chooses the next way using path-type movement model. The class
\texttt{PathMemoryMovement}\index{Class!PathMemoryMovement@\texttt{PathMemoryMovement}}
is defined in the file
\texttt{pathmemory.py}\index{Files!pathmemory.py@\texttt{pathmemory.py}}.

\section{Path Type with Restricted Movement}
\label{sec:pathrestricted}
\index{Path Type with Restricted Movement} Path Type with Restricted
Movement is designed for specific type of nodes where the movement
needs to be controlled through certain road-types. For the example
discussed at the beginning of Chapter~\ref{chp:approach}, two-wheeler
nodes are destined to travel through the remote roads to gather
information from sensors. Due to the structure of complex road
networks, the density of remote roads is higher compared to that of
highways. Therefore, there may occur instances that two-wheeler nodes
get trapped in remote roads once they enter into such roads,
especially if they follow random movement. Such cases reduce the
chances for a two-wheeler to meet with four-wheeler nodes, which
travel only through the highways. In restricted movement, when a
two-wheeler comes to a junction where a highway is connected, its
movement get restricted only through the highways until it meets with
a four-wheeler to transfer its data. Upon the data transfer, the
restriction is withdrawn and two-wheeler is allowed to continue its
journey through the remote roads. In other words, the restriction of
two-wheeler movement increases the probability of
two-wheeler-four-wheeler contact. Similar to Path Memory Movement,
Restricted Movement inherits the properties of Path Type Movement and
redefines the function
\texttt{compute\_next\_node()}\index{Functions!compute\_next\_node@\texttt{compute\_next\_node}}
with the addition of path restriction as shown in the following code.
\begin{lstlisting}[style=customstyle]
  # Addition of restriction av_nodes = [] # List for storing connected
  highways if self.buffer: # If object has data in its buffer for node
  in possible_nodes: e_type =
  road_graph[self.next_node][node]['e_type'] if e_type == 'highway':
  av_nodes.append(node) if av_nodes: possible_nodes = av_nodes[:] #
  End of restriction
\end{lstlisting}

In case of junctions which do not have a highway connected to them,
the node follows path-type based movement model. The restricted
movement is implemented as class
\texttt{RestrictedMovement}\index{Class!RestrictedMovement@\texttt{RestrictedMovement}}
in the file
\texttt{restricted.py}\index{Files!restricted.py@\texttt{restricted.py}}.

\section{Path Type with Wait Movement}
\label{sec:pathwait}
\index{Path Type with Wait Movement} Path Type with Wait is a
variation of Path Type with Restricted Movement on the fact that,
whenever a node reaches a particular junction, the node stops its
movement until the occurrence of a specific event. For example, a
two-wheeler comes with data to a junction where a highway is
connected, it waits at that junction for a four-wheeler instead of
roaming through the highways as the case with restricted
movement. Wait movement is implemented as a class
\texttt{WaitMovement}\index{Class!WaitMovement@\texttt{WaitMovement}},
derived from \texttt{PathTypeMovement}, in the file
\texttt{wait.py}. An additional attribute \texttt{wait\_flag} is
defined to control the waiting of a node at the junctions. The
function
\texttt{compute\_next\_node()}\index{Functions!compute\_next\_node@\texttt{compute\_next\_node}}
is redefined with the wait process as follows.
\begin{lstlisting}[style=customstyle]
  # Addition of Waiting Process Code av_nodes = [] if self.buffer: for
  node in possible_nodes: e_type =
  road_graph[self.next_node][node]['e_type'] if e_type == 'highway':
  av_nodes.append(node) if av_nodes: if not self.wait_flag:
  self.wait_start = global_params.sim_time self.wait_flag = True
  return self.next_node # End of Waiting Process Code
\end{lstlisting}

The value of \texttt{wait\_flag} becomes \texttt{True} when a
two-wheeler reaches a highway junction, i.e., the two-wheeler waits
when its \texttt{wait\_flag = True}. In our example, when a
four-wheeler comes in contact with the two-wheeler, the value of
\texttt{wait\_flag} turns to \texttt{False} and two-wheeler resumes
its journey as per the path type movement model. The value of
\texttt{wait\_flag} is assigned to \texttt{True} in the routing
protocol since the decision depends on the communication between
nodes.

\section{Addition of New Movement Models}\label{sec:mmadd}
Object oriented programming approach used in UDTN Simulator provides
an easy way to define custom movement models. The class
\texttt{SimpleRandomMovement} provides the basic movement for the
objects and \texttt{PathTypeMovement} provides a realistic movement of
vehicles through the road network. In order to incorporate a new
movement model, say
\texttt{ABCMovement}\index{Class!ABCMovement@\texttt{ABCMovement}}, we
need follow 3 steps.
\begin{description}
\item[Step 1:] Create a file \texttt{abc.py} in
  \texttt{movementmodel/} directory with the definition of movement
  model as a class. An example skeleton of \texttt{abc.py} is as
  follows: 
  \begin{lstlisting}[style=customstyle]
import random
import simplerandom as sr

class ABCMovement(sr.SimpleRandomMovement):
    # ABC movement model 
    
    ## Constructor 
    def __init__(self, obj_label, group_id, group_params,\ global_params):
        sr.SimpleRandomMovement.__init__(self, obj_label, group_id, group_params, global_params)

    ## Re-definition of initial_node for ABC movement
    def compute_initial_node(self, global_params):
        # Statement here
        
        return initial_node

    ## Redefinition of compute_next_node for ABC movement
    def compute_next_node(self, global_params):
        # Statements here

        return next_node
  \end{lstlisting}
\item[Step 2:] Edit the function \texttt{create\_mvmt\_object()} in
  \texttt{init/initsim.py} by adding an \texttt{elif} statement with
  information of \texttt{ABCMovement}. An example code is given below:
  \begin{lstlisting}[style=customstyle]
def create_mvmt_object(obj_label, group_id, group_params, global_params):
    obj_type = group_params['Movement']
    if obj_type == 'Stationary':
        return stationary.Stationary(obj_label,\ 
            group_id, group_params, global_params)
    elif obj_type == 'SimpleRandomMovement':
        return simplerandom.SimpleRandomMovement(obj_label, group_id, group_params, global_params)
    elif obj_type == 'PathTypeMovement':
        return pathtype.PathTypeMovement(obj_label,\
            group_id, group_params, global_params)
    elif obj_type == 'ABCMovement':
        return abc.ABCMovement(obj_label,\
            group_id, group_params, global_params)
  \end{lstlisting}
\item[Step 3:] Assign the movement model for a group of mobile nodes
  using the group parameter \texttt{Movement} in settings file
  \texttt{sim.config} as follows:
  \begin{lstlisting}[style=customstyle]
    Movement = ABCMovement
  \end{lstlisting}
\end{description}

Inheritance in object oriented programming helps us to create new
movement models by defining only the changes that is required from the
existing classes. Here, since the computation of \texttt{next\_node}
determines the movement, the redefinition of function
\texttt{compute\_next\_node()} acts as the core part of any mobility
model.

\section{Summary}
\label{sec:chp3summary}
In this chapter, we discussed different mobility models implemented in
UDTN Simulator and the procedure to add a new model. Along with
mobility models, the routing protocols play crucial role in
performance of any Delay Tolerant Network. Therefore, we discuss the
routing protocols in the next chapter.

\chapter{Routing Protocols}\label{chp:protocols}Similar to 
mobility models, routing protocols\index{Routing protocols} have a
crucial role in determining the performance of Delay Tolerant
Networks. Routing algorithms determine whether to send the data
whenever two nodes comes in contact with each other and, thereby, the
number of message copies in the network. Examples of routing
algorithms include epidemic protocol\index{Epidemic routing/protocol},
spray and wait, maxprop, and PROPHET. The objective of routing
algorithms is to achieve maximum message delivery ratio with minimum
number of message copies as well as within minimum time. In UDTNSim,
we implement three routing protocols ({\it i})~\emph{Epidemic
  Routing}, ({\it ii})~\emph{Superior Only Handoff}, and ({\it
  iii})~\emph{Superior Peer Handoff}\footnote{The source code for
  different routing protocols implemented in UDTNSim is available at\\
  \href{https://github.com/iist-sysnet/UDTNSim/tree/master/protocols}{\tt
    https://github.com/iist-sysnet/UDTNSim/tree/master/protocols}}.

\section{Epidemic Routing}\label{sec:epidemic}
Epidemic routing~\cite{vahdat2000epidemic} is the simplest protocol
for data handoffs in a DTN scenario. The protocol works as follows:
When an object~$A$ comes in contact with an object~$B$, $A$ sends its
messages (which are not in the buffer of~$B$) to~$B$. Even though
epidemic protocol has high delivery probability, more message copies
in the network result in high communication overhead and buffer
overflow. Therefore, different protocols are designed to control the
flooding of messages with the help of heuristics and optimization
parameters.

In UDTNSim, the routing algorithms are defined in \texttt{protocols/}
directory. Similar to movement model, we define each routing protocol
as a class and assign the protocol to an object using the group
attribute
\texttt{Protocol}\index{Parameters!Protocol@\texttt{Protocol}} in the
settings file
\texttt{sim.config}\index{Files!sim.config@\texttt{sim.config}}. An
object for routing protocol is also created for each movement object
and is assigned to the attribute \texttt{protocol\_obj} in the
respective movement
model\index{Variables!protocol\_obj@\texttt{protocol\_obj}}. We
require two attributes for a protocol object to handle the data
handoff process.
\begin{enumerate}
\item
  \texttt{neighbor\_dict:}\index{Variables!neighbor\_dict@\texttt{neighbor\_dict}}
  Dictionary to store the information of neighbors. That is, at any
  instant the dictionary contains the information of neighbor nodes to
  which the data can be transferred. The dictionary is of the form
  \begin{lstlisting}[style=customstyle]
    neighbor_dict = {n1: [d, f, TS], n2: [d, f, TS],...}.
  \end{lstlisting}
  Here, \texttt{n1} is the movement object identifier, \texttt{d} is
  the geographical distance toward \texttt{n1}, \texttt{f} is a flag
  to represent the continuity of contacts~(only used for displaying
  contacts in GUI), and \texttt{TS} is the time-stamp at which
  \texttt{n1} came into contact.
\item
  \texttt{contact\_objs:}\index{Variables!contact\_objs@\texttt{contact\_objs}}
  List to store the contact log of the current node. Each element in
  the list is a tuple of the form
  \begin{lstlisting}[style=customstyle]
    (n_label, TS_entry, TS_exit),
  \end{lstlisting}
  where \texttt{n\_label} is the unique identifier of the neighbor
  object, \texttt{TS\_entry} is the time-stamp at which the neighbor
  comes in contact, and \texttt{TS\_exit} is the time-stamp at which
  the neighbor comes out of the range of the current object.
\end{enumerate}

The routing protocol for each node consists of two steps: ({\it
  i})~find the neighbors and ({\it ii})~transfer the data to
neighbors. For performing the two steps, we define the following
functions:
\begin{enumerate}
\item
  \texttt{execute\_protocol():}\index{Functions!execute\_protocol@\texttt{execute\_protocol}}
  The method calls two functions, \texttt{find\_neighbors()} and
  \texttt{exchange \_data()}. For each movement object, this function
  is called from the function \texttt{execute \_simulation()} in
  \texttt{main.py} file.
  \begin{lstlisting}[style=customstyle]
    for mvmt_ob in mvmt_obj_list:
    mvmt_ob.protocol_obj.execute_protocol(mvmt_ob,\ global_params)
  \end{lstlisting}

  At each time-step, the position of object changes and the neighbor
  list gets recomputed.
\item
  \texttt{find\_neighbors():}\index{Functions!find\_neighbors@\texttt{find\_neighbors}}
  The function searches for the objects which are within the
  transmission range of the current object. The transmission range of
  nodes are taken from the group specific attribute \texttt{TX\_Range}
  from settings file \texttt{sim.config}. The object finds out the
  distance from other objects using \texttt{calculate\_distance()}
  function defined in \texttt{libs/geocalc.py} with the geographical
  position of two objects as arguments. If any of the nodes' distance
  is within \texttt{TX\_Range} of current object, its information
  added into the dictionary
  \texttt{neighbor\_dict}\index{Variables!neighbor\_dict@\texttt{neighbor\_dict}}.
  \begin{lstlisting}[style=customstyle]
    # If distance < transmission range both are in range if dist <=
    tx_range: if obj in self.neighbor_dict: self.neighbor_dict[obj][0]
    = dist self.neighbor_dict[obj][1] = True else: time_stamp =
    simtimer.convert_HMS(global_params.sim_time)
    self.neighbor_dict[obj] = [dist, False, time_stamp, None]
  \end{lstlisting}
  If an object is not in the transmission range and it is existing in
  the \texttt{neighbor\_dict}~(i.e., the object went out of range),
  its information is appended to the list \texttt{contact\_objs} with
  the exiting timestamp. Further, the object's information is removed
  from \texttt{neighbor\_dict}.
  \begin{lstlisting}[style=customstyle]
    self.contact_objs.append((obj.obj_id, obj_infn[2],\ time_stamp))
    self.neighbor_dict.pop(obj)
  \end{lstlisting}

\item
  \texttt{exchange\_data():}\index{Functions!exchange\_data@\texttt{exchange\_data}}
  The actual data transfer is simulated in this function. The messages
  present in the buffer of current object as well as in the neighbor
  object are taken into lists \texttt{src\_msg\_data} and
  \texttt{dst\_msg\_data}, respectively. The messages that are not in
  neighbor object is found out with the statement
  \begin{lstlisting}[style=customstyle]
    msgs_not_in_node = set(src_msg_data)-set(dst_msg_data).
  \end{lstlisting}
  Each message in \texttt{msgs\_not\_in\_node} is appended to the
  buffer of neighbor objects with the current object's identifier as
  the sender information.
  \begin{lstlisting}[style=customstyle]
    for msg in msgs_not_in_node: node.buffer.append([msg,
    mvmt_ob.obj_id])
  \end{lstlisting}
\item
  \texttt{print\_neighbors():}\index{Functions!print\_neighbors@\texttt{print\_neighbors}}
  This function displays information in GUI when nodes comes in
  contact with each other. It avoids the continuous display of
  neighbor nodes using the flag in the \texttt{neighbor\_dict}.
\end{enumerate}

\section{Superior Only Handoff Protocol}
\label{sec:superioronly}
The major drawback of epidemic protocol is its high communication
overhead. Therefore, different protocols are designed aiming at
reducing the communication overhead while maintaining high delivery
ratio. Superior Only Handoff~\cite{jain2014disaster} is designed for
scenarios where the nodes are classified in a hierarchical
structure. Considering the example discussed in
Chapter~\ref{chp:approach}, the scenario consists of a geographical
area with different types of roads (such as remote roads and highways)
as well as different types of vehicles (two-wheeler, four-wheeler, and
pedestrians). In such a case, the nodes can be classified into
different levels according to the types of roads that they can
travel. For instance, a 3-level classification is possible, i.e.,
four-wheeler at Level-0 (access only to highways), two-wheeler at
Level-1, i.e., as children of four-wheeler (access to highways and
remote roads), and pedestrian at the lowest level (access to all types
of roads), i.e., Level-2. The nodes at Level-0 (four-wheeler nodes)
are superior to nodes residing Level-1 (two-wheeler) as well as
Level-2 (pedestrians). Nodes at Level-1 is superior only to Level-2
nodes. Therefore, in Superior Only handoff, a node transfers its data
only to nodes which are superior in the classification hierarchy. In
other words, a two-wheeler can transfer only to a four-wheeler, not to
a two-wheeler or a pedestrian.

Similar to movement models, Superior Only Handoff protocol is defined
as a class \texttt{Superior
  OnlyHandoff}\index{Class!SuperiorOnlyHandoff@\texttt{SuperiorOnlyHandoff}}
in the file
\texttt{superioronly.py}\index{Files!superioronly.py@\texttt{superioronly.py}}. The
class derive properties from
\texttt{EpidemicHandoff}\index{Classes!EpidemicHandoff@\texttt{EpidemicHandoff}}
and override the function
\texttt{exchange\_data()}\index{Functions!exchange\_data@\texttt{exchange\_data}}
with code to discard the neighbors which are at equal or lower level
in the hierarchy.
\begin{lstlisting}[style=customstyle]
for node in self.neighbor_dict:
    # Discard the nodes at equal or lower levels
    if self.level <= node.protocol_obj.level:
      continue
    # Code to exchange data to neighbors
\end{lstlisting}
\vspace{-3.5mm}\hspace{2cm}\vdots

\noindent In addition to the function \texttt{exchange\_data()}, the
class has an additional attribute
\texttt{level}\index{level@\texttt{level}} which assigns a level to
the object in the hierarchy. In order to find the level of an object,
a separate function
\texttt{find\_level()}\index{Functions!find\_level@\texttt{find\_level}} is
defined and is invoked from the constructor
\texttt{\_\_init\_\_()}. The level of an object can be computed as
\[
  \textit{Level} = \textit{Maximum Path\_Type available} -
  \textit{Minimum Path\_Type the object can travel}.
\]
The function \texttt{find\_level()} makes use of general parameters
and group specific parameters (discussed in
Section~\ref{subsec:simparams}) specified in configuration file
\texttt{sim.config} to compute an object's level.

\section{Superior Peer Handoff Protocol}
\label{sec:superiorpeer}
\index{Superior Peer Handoff} Since superior only handoff restricts
the communication between nodes at the same level, the protocol may
reduce the probability of data reaching the nodes at higher
level. However, the presence of message copies at multiple nodes at a
level increases the chance of message to reach nodes at higher
levels. Therefore, Superior Peer Handoff~\cite{jain2014disaster}
allows a node to handover its data not only to its superior, but to
its peers (i.e., nodes at the same level). Here, a two-wheeler is
allowed to transfer its data to another two-wheeler unlike Superior
Only Handoff. The protocol is defined as a class
\texttt{SuperiorPeerHandoff}\index{Class!SuperiorPeerHandoff@\texttt{SuperiorPeerHandoff}}
derived from
\texttt{SuperiorOnlyHandoff}\index{Class!SuperiorOnlyHandoff@\texttt{SuperiorOnlyHandoff}}
with a change in the selection of neighbors as follows:
\begin{lstlisting}[style=customstyle]
for node in self.neighbor_dict:
    # Discard the nodes at lower levels
    if self.level < node.protocol_obj.level:
      continue
    # Code to exchange data to neighbors
\end{lstlisting}
\vspace{-3.5mm}\hspace{2cm}\vdots

\noindent The class \texttt{SuperiorPeerHandoff} is defined in the
file
\texttt{superiorpeer.py}\index{Files!superiorpeer.py@\texttt{superiorpeer.py}}.

\section{Addition of New Routing Protocols}\label{sec:protoadd}
Similar to movement models, for the addition of a new routing
protocol, say
\texttt{TestRouting}\index{Class!TestRouting@\texttt{TestRouting}}, we need
to follow three steps.
\begin{description}
\item[Step 1:] Create a file \texttt{test.py} in the directory
  \texttt{protocols/} with the class definition of new protocol. Since
  neighborhood discovery process for all protocols is the same, the
  custom protocol can inherit it from epidemic protocol. The
  difference comes only in deciding the neighbor to which data is
  transferred to, i.e., the definition of function
  \texttt{exchange\_data()}\index{Functions!exchange\_data@\texttt{exchange\_data}}. A
  sample skeleton for a custom data routing protocol, say
  \texttt{TestRouting}\index{Class!TestRouting@\texttt{TestRouting}} is as
  follows: 
  \begin{lstlisting}[style=customstyle]
    ''' module of Test data routing '''
import geocalc
import simtimer
import epidemic as ep

class TestRouting(ep.EpidemicRouting):
    ''' Test Data Routing '''

    def __init__(self):
        ep.EpidemicRouting.__init__(self)

    ## Function for exchanging data with neighbors
    def exchange_data(self, mvmt_ob, global_params):
        # Redefinition of exchange data
        
        return
  \end{lstlisting}
\item[Step 2:] Edit the function \texttt{create\_routing\_object()} with an
  addition of an \texttt{elif} statement for \texttt{TestRouting} as
  follows: 
  \begin{lstlisting}[style=customstyle]
    def create_handoff_object(group_params):
    protocol_type = group_params['Protocol']
    if protocol_type == 'EpidemicProtocol':
        return epidemic.EpidemicProtocol()
    elif protocol_type == 'TestProtocol':
        return test.TestProtocol()

    return None
  \end{lstlisting}
\item[Step 3:] Assign routing protocols to the nodes by assigning the new
  protocol name to the group parameter \texttt{Protocol} in the
  settings file \texttt{sim.config}.
  \begin{lstlisting}[style=customstyle]
    Protocol = TestRouting
  \end{lstlisting}
\end{description}

Similar to movement models, property of inheritance allows us to
define new routing protocol with minimum effort. That is, the major
change is the redefinition of \texttt{exchange\_data()} function.

\section{Summary}
\label{sec:chp4summary}

In this chapter, we discussed different protocols required for routing
the data from source to destination. The UDTN simulator implements
three routing protocols, ({\it i})~Epidemic Routing, ({\it
  ii})~Superior Only Handoff, and ({\it iii})~Superior Peer
Handoff. The supporting libraries for movement models and routing
protocols are discussed in next chapter.

\chapter{Supporting Libraries}\label{chp:supplib}
Apart from the major part of the simulator, i.e., the mobility models
and routing protocols, we define some additional functions to support
the operations of the simulator. The additional libraries include
Graphical User Interface~(GUI), events, and reports.

\section{Graphical User Interface}\label{sec:gui}
\begin{figure}[h]
  \centering
  \includegraphics[width=0.9\textwidth]{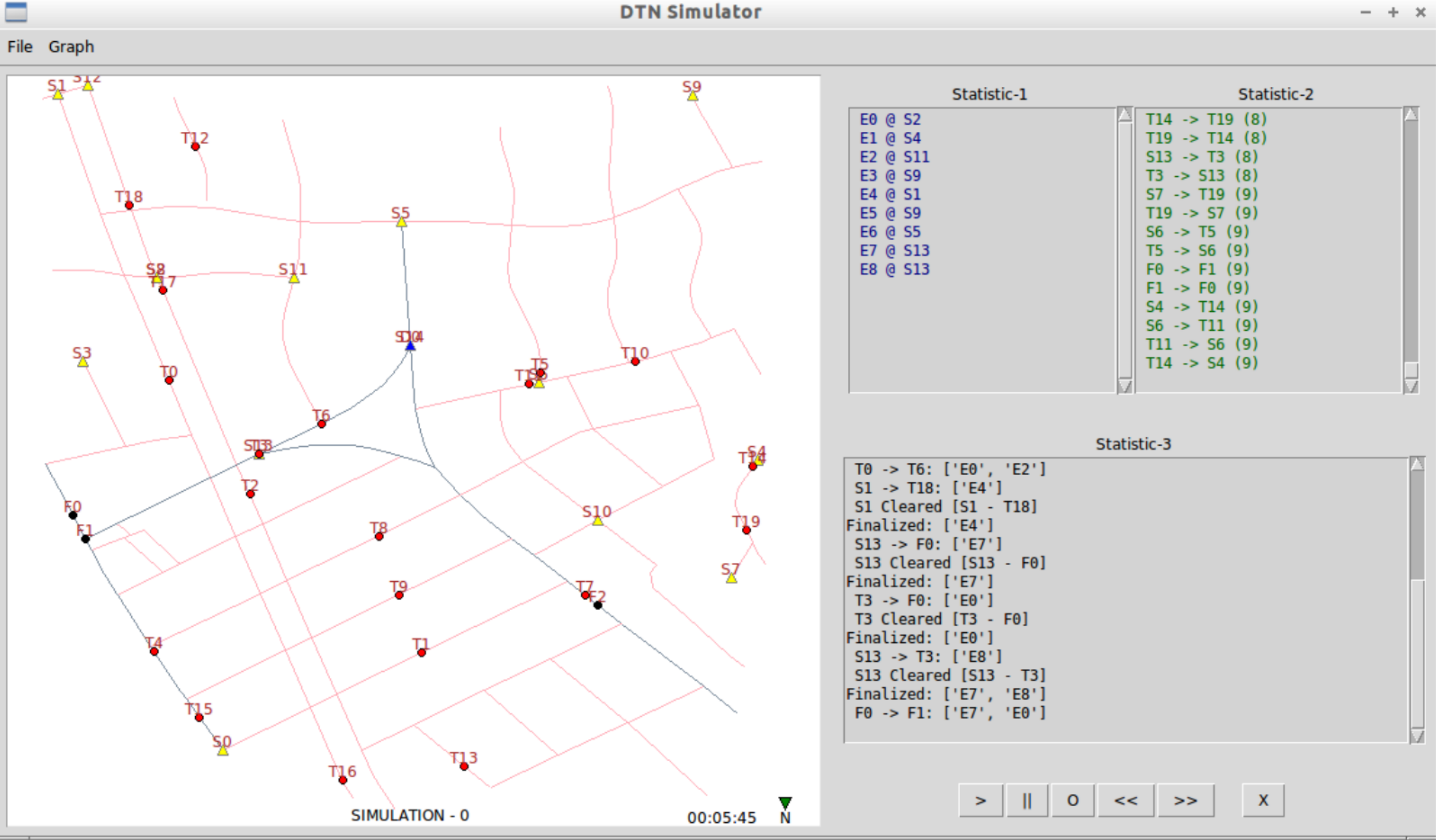}
  \caption{Snapshot of simulator GUI.}
  \label{fig:simulator}
\end{figure}

Graphical User Interface\index{Graphical User Interface~(GUI)} is
created using Python
\emph{Tkinter}\cite{tkinter}\index{Tkinter@\textit{Tkinter}}
library\footnote{The source code for GUI module is available at\\
  \href{https://github.com/iist-sysnet/UDTNSim/tree/master/gui}{\tt
    https://github.com/iist-sysnet/UDTNSim/tree/master/gui}}. The
simulator GUI can be enabled using the value of environment parameter
\texttt{GUI\_Enabled = True/False} in the settings file
\texttt{sim.config}. GUI is defined as a class
\texttt{Gui}\index{Class!Gui@\texttt{Gui}} in \texttt{gui/gui.py}
file. A canvas, with dimension \mbox{$650\times 640$} pixels, is
defined to display the map and movement of mobile nodes. A screenshot
of the simulator is shown in Figure~\ref{fig:simulator}.

Buttons are provided at the bottom right for controlling the
simulation. The buttons include play~($>$), pause~($||$), stop~(o),
decelerate~($<<$), accelerate~($>>$), and exit~(X). Further, three text
boxes are provided to display the status of simulation. The messages
can be displayed in each of them using the function
\texttt{print\_msg()}. The functions required for GUI are defined as
follows:
\begin{enumerate}
\item
  \texttt{draw\_canvas():}\index{Functions!draw\_canvas@\texttt{draw\_canvas}}
  This function draws the map and mobile nodes within the canvas using
  the functions \texttt{create\_map()} and \texttt{create\_node()},
  respectively.
\item \texttt{create\_map():}\index{Functions!create\_map@\texttt{create\_map}}
  The function iterates through each way in the
  \texttt{way\_dict}\index{Variables!way\_dict@\texttt{way\_dict}} and draws a
  straight line using the end points of the way. The coordinates for
  endpoints is taken from the
  \texttt{node\_dict}\index{Variables!node\_dict@\texttt{node\_dict}}.
\item
  \texttt{create\_node():}\index{Functions!create\_node@\texttt{create\_node}}
  The mobile objects are displayed as filled circles with the color
  specified in the group parameter \texttt{Color} in the settings file
  \texttt{sim.config}. Static nodes are created as filled
  triangles. Each mobile object is associated with a label
  corresponding to its unique identifier.
\item
  \texttt{redraw\_node():}\index{Functions!redraw\_node@\texttt{redraw\_node}}
  After each shift in the position of a mobile node, it is redrawn to
  the new position using the function \texttt{redraw\_node()}. The
  function is called from the
  \texttt{execute\_simulation()}\index{Functions!execute\_simulation@\texttt{execute\_simulation}}
  in \texttt{main.py} file, just after the
  \texttt{update\_position()}\index{Functions!update\_position@\texttt{update\_position}}
  function.
  \begin{lstlisting}[style=customstyle]
    # A unit movement of mobile node
    mvmt_obj.update_position(global_params) # Redraw the node's
    position if global_params.gui_ob:
    global_params.gui_ob.redraw_node(mvmt_obj, \ global_params)
  \end{lstlisting}
\item \texttt{update\_sim():}\index{Functions!update\_sim@\texttt{update\_sim}}
  Upon estimating the position of mobile nodes evaluated, the
  simulator GUI is updated using the
  \texttt{update\_sim()}function. It is called from the
  \texttt{execute \_simulation()} function in \texttt{main.py}.
  \begin{lstlisting}[style=customstyle]
    # Update simultion parameters
    global_params.gui_ob.update_sim(global_params)
  \end{lstlisting}
  In this function, the canvas is redrawn according to the simulation
  speed. Also, the simulation timer is updated using
  \texttt{update\_sim\_timer()} function.
\item \texttt{print\_msg():}\index{Functions!print\_msg@\texttt{print\_msg}}
  Function used to display messages from simulation. Three text
  boxes (Statistics-1, 2, and 3) are provided to display the
  simulation messages. Messages to three boxes can be displayed by
  changing the first argument of the function. The syntax of the
  function is as follows:
  \begin{lstlisting}[style=customstyle]
    print_msg(msg_type, msg)
  \end{lstlisting}
  Here, \texttt{msg\_type} has values \texttt{STAT1, STAT2, }and
  \texttt{STAT3}, for three text boxes respectively. The second
  argument is the message to be displayed.
\end{enumerate}
GUI-related functions are called using the object created for GUI. In
simulator the object can be accessed using
\texttt{global\_params.gui\_ob}.

\section{Events}\label{sec:events} Event\index{Events} is a term that
represents the occurrence of an action. In our context, an event
includes an emergency call~(in case of a disaster situation) and
message generation in a sensor~(in case of data gathering). We define
the classes and functions required for events in \texttt{event/}
directory\footnote{The source code for event operations are available
  at \\
  \href{https://github.com/iist-sysnet/UDTNSim/tree/master/event}{\tt
    https://github.com/iist-sysnet/UDTNSim/tree/master/event}}.

\subsection{Event}\label{subsec:event} We define an \emph{event} using a
class \texttt{Event}\index{Class!Event@\texttt{Event}} defined in
\texttt{event/events.py}\index{Files!events.py@\texttt{events.py}}. The
attributes of \texttt{Event} class are as follows:
\begin{enumerate}
\item
  \texttt{event\_interval\_list:}\index{Variables!event\_interval\_list@\texttt{event\_interval\_list}}
  The rate of events are specified in the settings file
  \texttt{sim.config} using the parameter
  \texttt{Random\_Msg\_Gen\_Parameter}\index{Parameters!Random\_Msg\_Gen\_Parameter@\texttt{Random\_Msg\_Gen\_Parameter}}. Further,
  the time instants for which events need to be generated is computed
  and appended to \texttt{event\_interval\_list}. The list is
  populated by calling a function \texttt{init\_events()} defined in
  \texttt{init/initsim.py}.
\item \texttt{event\_counter:} Class variable which iterates through
  the \texttt{event\_interval\_list}.
\item
  \texttt{stat\_obj\_list:}\index{Variables!stat\_obj\_list@\texttt{stat\_obj\_list}}
  Class variable, a subset of \texttt{mvmt\_obj\_list}, with a list of
  stationary objects~(sensors and stationary routers) in the
  simulation environment. An event is created and assigned to one of
  the objects chosen randomly from this stationary nodes. Here, we
  assume that events are generated only in stationary objects).
\item \texttt{e\_id:} Unique event identifier~(For example E1, E2, and
  so on.)
\item \texttt{time:} Event occurrence time.
\item \texttt{duration:} The time at which the event lasts.
\item \texttt{data:} Data generated in the event.
\item \texttt{expiry:} Time of event expiry~(\texttt{time +
    duration}).
\item \texttt{expired\_status:} It is a boolean variable. If the event
  is expired, its value will be \texttt{True} and \texttt{False},
  otherwise. When an event is generated, its value is set to
  \texttt{False}. The function \texttt{check\_expiry()} continuously
  check for the expiration of the event.
\item \texttt{buffer:} List to store miscellaneous information of the
  event.
\item \texttt{report\_obj:} Report object to create logs of the
  event~(Section~\ref{sec:reports}).
\end{enumerate}

\noindent The member function of the class \texttt{Event} includes
\texttt{check\_expiry()}.
\begin{enumerate}
\item
  \texttt{check\_expiry():}\index{Functions!check\_expiry@\texttt{check\_expiry}}
  This function continuously checks for the expiration time of each
  event. If the simulation time exceeds expiration time, the attribute
  \texttt{expired\_status} is set to \texttt{True}. The function is
  called for each event from the function
  \texttt{execute\_simulation()} in \texttt{main.py}.
  \begin{lstlisting}[style=customstyle]
    for event_ob in global_params.events_list: if not
    event_ob.expired_status:
    event_ob.check_expiry(global_params.sim_time)
  \end{lstlisting}
\end{enumerate}

\subsection{Event Operations}\label{subsec:eventsops}
Since creation of events includes three operations, ({\it i})~creation
of \texttt{Event} objects, ({\it ii})~selection of stationary node on
which the event is assigned, and ({\it iii})~random message creation
for the event, we define a function
\texttt{create\_event()}\index{Functions!create\_event@\texttt{create\_event}}
in file
\texttt{event/eventops.py}\index{Files!eventops.py@\texttt{eventops.py}}. Event
message (by default, the length is 5~bytes) is created using a
function \texttt{create\_random\_data()}. The identifier of selected
movement object is appended to the event's buffer. To the
static/mobile node's buffer, a tuple of the form \texttt{(event\_ob,
  prev\_obj\_id)} is appended, where
\texttt{event\_ob}\index{Variables!event\_ob@\texttt{event\_ob}} is the event
object and \texttt{prev\_obj\_id} is the identifier of the
static/mobile node which handed over the event. The created event
object is appended to the global variable
\texttt{global\_params.events\_list}\index{Variables!events\_list@\texttt{events\_list}}.

\section{Reports}\label{sec:reports}\index{Reports}
Logs are important parts of any simulation. Creating logs in a format
convenient to the user is of much importance in data
analytics. Therefore, we define separate classes for creating
reports\footnote{The source code for classes defined for Reports
  module is available at \\
  \href{https://github.com/iist-sysnet/UDTNSim/tree/master/report}{\tt
    https://github.com/iist-sysnet/UDTNSim/tree/master/report}}. The
definitions are given in
\texttt{report/report.py}\index{Files!report.py@\texttt{report.py}}
file. A function
\texttt{create\_report\_directory()}\index{Functions!create\_report\_directory@\texttt{create\_report\_directory}}
is defined in \texttt{report.py} to create a log directory given using
parameter \texttt{Report\_Directory} in settings file
\texttt{sim.config}\index{Files!sim.config@\texttt{sim.config}}. The
function is called while initializing the simulation environment in
\texttt{init\_sim\_envt()}\index{Functions!init\_sim\_envt@\texttt{init\_sim\_envt}}
in
\texttt{init/initsim.py}\index{Files!initsim.py@\texttt{initsim.py}}.
\begin{lstlisting}[style=customstyle]
  # Create a report directory report_status =
  reports.create_report_directory(\ global_params)
\end{lstlisting}

\noindent We define report classes separately for movement objects as
well as for the events.

\subsection{Movement Report}\label{subsec:movementreport}Each object,
static and mobile, needs to create logs according to the simulation
requirements. Therefore, we define the class
\texttt{MovementReport}\index{Class!MovementReport@\texttt{MovementReport}}
in \texttt{report.py}. When a movement object is created, an object of
\texttt{MovementReport} class is also created and assigned to the
attribute \texttt{report\_obj}\index{Variables!report\_obj@\texttt{report\_obj}}
of the movement object. Using \texttt{report\_obj} each movement
object can call functions in this class. The class is made with
sufficient flexibility that, we can add more functions or we can
include the code to create our own log in the function
\texttt{create\_log()}. The constructor~(\texttt{\_\_init\_\_()})
creates a subdirectory~(object identifier \texttt{obj\_id} as the
directory name) for each movement object for recording the logs
specific to the object.

\begin{enumerate}
\item \texttt{create\_log():}\index{Functions!create\_log@\texttt{create\_log}}
  The function creates different files describing the activities of
  the mobile node. For example, the file \texttt{contacts\_0.dat}
  contains the details of contact with other nodes with neighbor
  identifier, contact starting time, and contact ending time in
  simulation 0. The information can be directly obtained from the
  attribute \texttt{protocol\_obj.contact \_objs} of each
  object. Similarly, other files are created for information on
  message collection, ways traversed, and summary of the movement. The
  function is called as the final step of each simulation, i.e., from
  the function
  \texttt{execute\_simulation()}\index{Functions!execute\_simulation@\texttt{execute\_simulation}}
  in \texttt{main.py}\index{Files!main.py@\texttt{main.py}}.
  \begin{lstlisting}[style=customstyle]
    # Generating the log of the nodes' movement in the simulation for
    mvmt_ob in global_params.mvmt_obj_list: mobility_status =
    global_params.envt_params.groups\ [mvmt_ob.group_id]['Mobile'] if
    not mobility_status: continue else:
    mvmt_ob.report_obj.create_log(mvmt_ob, \ global_params)
  \end{lstlisting}
\end{enumerate}

\subsection{Event Report}\label{subsec:eventreport} For creating logs
for events, we define a class \texttt{EventReport} in the file
\texttt{report/report.py}. A subdirectory \texttt{events} will be
created by the constructor. The \texttt{create\_log()} function has
the code to write the information related to the events.
\begin{enumerate}
\item \texttt{create\_log():}\index{Functions!create\_log@\texttt{create\_log}}
  Similar to the function in \texttt{MovementReport}, this function
  writes information of an event to a file in \texttt{events}
  directory. For each event, the function creates a file with name
  same as the event identifier~(\texttt{e\_id}) concatenated with the
  simulation number at the end of filename. The attributes such as
  event id, time of origin, time of expiry, expired status, data
  related to the event, and its buffer information are written into
  the file. The function is called from \texttt{execute\_simulation()}
  in \texttt{main.py} after each simulation.
  \begin{lstlisting}[style=customstyle]
    # Generating the log of the events for event_ob in
    global_params.events_list:
    event_ob.report_obj.create_log(event_ob)
  \end{lstlisting}
\end{enumerate}

\section{Additional Libraries}\label{sec:additionallibs} We define
some additional functions required for different components of the
simulator in \texttt{libs/} directory\footnote{The source code for
  additional libraries is available at\\
  \href{https://github.com/iist-sysnet/UDTNSim/tree/master/libs}{\tt
    https://github.com/iist-sysnet/UDTNSim/tree/master/libs}}.
\begin{enumerate}
\item
  \texttt{simtimer.py:}\index{Files!simtimer.py@\texttt{simtimer.py}}
  The function creates a timer for the simulation. That is, the real
  time~(i.e., each clock tick of simulator timer) corresponding to a
  unit
  \texttt{update\_position()}\index{Functions!update\_position@\texttt{update\_position}}
  function call. Simulator timer depends on the map and maximum speed
  of the mobile objects. The computed clock tick is added to the
  simulator clock after each \texttt{update\_position()} from
  \texttt{execute\_simulation()} in \texttt{main.py}.
  \begin{lstlisting}[style=customstyle]
    # Updating the simulation clock timer global_params.sim_time +=
    global_params.sim_tick
  \end{lstlisting}

  The simulator clock is initialized from \texttt{init\_sim\_envt()}
  in \texttt{init/initsim.py}. It also creates the geographical
  distance~(\texttt{pix\_unit}) corresponding to a pixel
  unit. \texttt{pix\_multiplier} is a dictionary with values for each
  group of mobile objects. The value is used for interpolating the
  points while creating the movement points in the function
  \texttt{populate\_way \_points()} in \texttt{simplerandom.py}. Each
  pixel position shift depends on the speed of the corresponding
  object.

  The function,
  \texttt{convert\_HMS()}\index{Functions!convert\_HMS@\texttt{convert\_HMS}}
  converts the time in hours to a string in HH:MM:SS format.

\item \texttt{geocalc.py:}\index{Files!geocalc.py@\texttt{geocalc.py}}
  The library consists of calculations related to geographical
  coordinates. The functions are as follows:
  \begin{enumerate}
  \item
    \texttt{calculate\_distance():}\index{Functions!calculate\_distance@\texttt{calculate\_distance}}
    It calculates the geographical distance between two points. The
    points are given as tuples with \mbox{(\texttt{lat, lon})} format.
  \item
    \texttt{geo\_to\_cart():}\index{Functions!geo\_to\_cart@\texttt{geo\_to\_cart}}
    Convert a geographical coordinate to the corresponding cartesian
    coordinate for the GUI.
  \item
    \texttt{cart\_to\_geo():}\index{Functions!cart\_to\_geo@\texttt{cart\_to\_geo}}
    Convert a cartesian coordinate in the GUI to its corresponding
    geographical coordinate.
  \end{enumerate}
\item \texttt{guicalc.py:}\index{Files!guicalc.py@\texttt{guicalc.py}}
  The computations required for GUI are implemented as functions in
  this file.
  \begin{enumerate}
  \item \texttt{translation\_factor():} Given a map, in order to
    accommodate it in the GUI canvas~(with cartesian coordinates
    \texttt{(x,y)}, \texttt{x,y$\ge 0$}), we need to do translation
    and scaling operation. Given the upper and lower geographical
    bounds of the map, this function returns the translation factor
    for the map in geographical coordinates.
  \item \texttt{scale\_factor():} Given the bounds and the dimensions
    of the GUI canvas, the function returns the scaling factor
    corresponding to latitude and longitude.
  \item \texttt{compute\_delay\_pixels():} Returns a dictionary with
    values for each of the group of nodes. In settings file
    \texttt{sim.config}, the delay experienced by a node at a junction
    is given using the group parameter \texttt{Junction\_Delay}. While
    populating the movement points, we add dummy geographical points
    at the end of the road to create the delay. The number of dummy
    points required depends on the speed of node as well as the value
    of \texttt{pix\_unit}.
  \end{enumerate}
\item
  \texttt{graphops.py:}\index{Files!graphops.py@\texttt{graphops.py}}
  Defines the function for graph operations for the road network.
  \begin{enumerate}
  \item \texttt{create\_minimal\_graph():} The function creates a
    graph, using Python
    \texttt{NetworkX}~\cite{networkx}\index{NetworkX@\textit{NetworkX}}
    library, from the dictionary \texttt{way\_dict}.
  \end{enumerate}
\item \texttt{shared.py:}\index{Files!shared.py@\texttt{shared.py}} We
  create two classes for the purpose of global variables since some
  simulation parameters need to be accessed in major components of the
  simulator. Therefore, we create a single point of entry for such
  attributes using these classes.
  \begin{enumerate}
  \item \texttt{Shared:}\index{Class!Shared@\texttt{Shared}} Defines a
    class with following attributes
    \begin{itemize}
    \item
      \texttt{envt\_params:}\index{Variables!envt\_params@\texttt{envt\_params}}
      It is a dictionary with general parameters and group-specific
      parameters (See Section~\ref{subsec:simparams}).
    \item
      \texttt{way\_dict:}\index{Variables!way\_dict@\texttt{way\_dict}}
      Dictionary with way information present in the map (See
      Section~\ref{subsec:xmlparser}).
    \item
      \texttt{node\_dict:}\index{Variables!node\_dict@\texttt{node\_dict}}
      Dictionary with information of geographical points extracted
      from the map (See Section~\ref{subsec:xmlparser}).
    \item \texttt{gui\_ob:}\index{Variables!gui\_ob@\texttt{gui\_ob}}
      The object of Gui class which represents the graphical user
      interface of the UDTN simulator.
    \item
      \texttt{gui\_params:}\index{Variables!gui\_params@\texttt{gui\_params}}
      Dictionary with parameters related to GUI. Its keys include
      \texttt{`pix\_ unit', `pix\_multiplier',} and
      \texttt{`delay\_pixels'}.
    \item
      \texttt{mvmt\_obj\_list:}\index{Variables!mvmt\_obj\_list@\texttt{mvmt\_obj\_list}}
      A list consists of objects created using movement models
      specified in the configuration file \texttt{sim.config}
    \item
      \texttt{road\_graph:}\index{Variables!road\_graph@\texttt{road\_graph}}
      Graph object corresponding to the road network. It is created
      using the function \texttt{create\_minimal\_graph()}. The
      library used for graph operations is
      \emph{networkx}~\cite{networkx}. The vertex ID's of
      \texttt{road\_graph} are the \texttt{node\_id}s representing
      end-points of ways. An edge represents a road connecting two
      vertices. Each edge is associated with three attributes ({\it
        i})~\texttt{weight}: represents the length of the road, ({\it
        ii})~\texttt{e\_type}: represents the type of road, and ({\it
        iii})~\texttt{way\_id}: represents the ID of the way, i.e.,
      key from \texttt{way\_dict}. The variable \texttt{road\_graph}
      simplifies access to map information while designing movement
      models.
    \item
      \texttt{sim\_time:}\index{Variables!sim\_time@\texttt{sim\_time}}
      Simulator clock in hours.
    \item
      \texttt{sim\_tick:}\index{Variables!sim\_tick@\texttt{sim\_tick}}
      The value of each clock-tick in hours.
    \item
      \texttt{writer\_obj:}\index{Variables!writer\_obj\texttt{writer\_obj}}
      An object of class \texttt{Colors} defined in
      \texttt{gui/textformat.py}. The class has a function
      \texttt{print\_msg()} to print text in colors.
    \item
      \texttt{events\_list:}\index{Variables!events\_list@\texttt{events\_list}}
      List of event objects created.
    \end{itemize}
    We create an object \texttt{global\_params} of class
    \texttt{Shared} as the first step of initialization process in
    \texttt{init\_sim\_event()}\index{Functions!init\_sim\_event@\texttt{init\_sim\_event}}.
    \begin{lstlisting}[style=customstyle]
      # Create a shared object for sharing global information
      global_params = shared.Shared()
    \end{lstlisting}
    each of these values are assigned using the return values of
    initialization functions. Further, we pass \texttt{global\_params}
    as the argument to functions as a single point entry for
    simulation parameters.
  \item \texttt{Controls:}\index{Class!Controls@\texttt{Controls}}
    Defines four control variables for the simulator GUI. It enables
    the programmer to check whether the user has set the simulator to
    pause mode or stop mode. Also, it has a counter
    \texttt{no\_of\_simulation} to track the count of simulations.
  \end{enumerate}
\end{enumerate}

\section{Summary}
\label{sec:chp5summary}
In this chapter, we discussed the essential libraries required for the
simulator to work. It includes the graphical user interface, reports,
event generation, timer, and associated operations. 

\chapter{Summary}\label{chp:summary}
Delay Tolerant Networking~(DTN)\index{DTN} is an emerging paradigm in
the field of networking which tries to address problems due to network
delays and disruptions. In the context of terrestrial environments the
disruptions occur mainly due to node mobility. Therefore, we developed
an Urban Delay Tolerant Network Simulator~(UDTNSim)\footnote{The
  source code of UDTNSim is available at
  \href{https://github.com/iist-sysnet/UDTNSim}{\tt
    https://github.com/iist-sysnet/UDTNSim}} using Python to study the
performance of systems involving mobile models following different
mobility models and routing protocols.

We use a modular approach for the simulator to simplify the
development and testing of new mobility models and protocols. The
modules include \emph{settings, initialization, parser, movement
  models, routing protocols, GUI,} and \emph{reports}. Development of
movement models and routing protocols are carried out using object
oriented programming approach to reduce the effort required in
implementing the models. In order to test mobility and protocols
visually, we added a graphical user interface with road
network~(distinguishing different types of roads) and mobile
nodes. GUI has provisions to display required information related to
simulations. A flexible report module is provided to create simulation
related logs.

\section{Related Work in UDTNSim}
\label{sec:relatedwork}
Urban DTN simulator is used primarily to study the data gathering
process in case of emergency situations such as disasters. The main
aim of our study is to design an efficient emergency response system
using sensor-based information gathering. We used agent based approach
for data collection, where two-wheeler, four-wheeler, and pedestrian
nodes are deployed as agents. A shanty town Dharavi, Mumbai, India is
chosen as the geographical region due to its complex road network. Two
mobility models, \textit{Path Type Based Model} and \emph{Path Memory
  Based Model} (as discussed in Chapter~\ref{chp:movementmodels}) are
designed with the assumption that the map information is not available
to the agents. Therefore, both the movement models use decisions based
on random choices at the junctions. Along with two movement models, we
developed three routing protocols ({\it i})~\emph{No Handoff}, ({\it
  ii})~\emph{Superior Only Handoff}, and ({\it iii})~\emph{Superior
  Peer Handoff} (See
Chapter~\ref{chp:protocols})~\cite{jain2014disaster}. With the
proposed movement models and routing protocols, we are able to achieve
a delivery ratio of only~20\%. This necessitates the need for further
research into the field of information gathering.

A study on dynamic path rescheduling model for mobile data collectors
is carried out with the assumption of prior map information. Here, we
consider two types of agents, two-wheeler and four-wheeler nodes, and
their mobility is decided based on the generation of random events in
the area. Mobile agents are assumed to have a direct communication
with a Command and Control Center, which decides the route for the
agents according to the events. Two mobility models ({\it
  i})~\emph{Minimum Deviated Walk}~(MW) and ({\it ii})~\emph{Ortho
  Walk}~(OW)~\cite{raj2015efficient} are designed for agents to reach
the event location within the expiration time. Our results show that
that availability of map information has significant influence on the
delivery ratio, i.e., both~MW and~OW provides more than~90\% of
delivery ratio.

Urban DTN simulator offers an easy way to develop and test mobility
models and routing protocols for mobile ad-hoc networks, especially
the network uses store-and-forward approach. As far as the simulator
is concerned, the modules can be further improved with more realistic
features by taking into account the characteristics of road networks
as well as vehicles. Also, the simulator can be extended to simulate
the characteristics of transport, network, and data link layers in
addition to the existing bundle layer-based architecture.

\phantomsection
\addcontentsline{toc}{chapter}{Bibliography}
\bibliographystyle{IEEEtran}
\bibliography{book}

\phantomsection
\addcontentsline{toc}{chapter}{Index}
\printindex

\end{document}